\documentstyle[12pt]{article}

\begin{document}
\begin{center}
{\bf \Large  Energy Loss in Nuclear Drell-Yan Process}

Jian-Jun Yang$^{1,2}$, Guang-Lie Li$^{2,3}$

1. Department of Physics, Nanjing Normal University, 
   Nanjing 210024,  P.R. China

2. Institute of High Energy Physics,
   P.O. Box 918(4), Beijing 100039, P.R. China

3. CCAST(World Laboratory), B.O. Box 8730, Beijing 100080, P.R. China  
  
\end{center}

\begin{abstract}

By means of the nuclear parton distributions which can be used
to provide a good explanation for the EMC effect in  the whole x range, 
we investigate the energy loss effect in nuclear Drell-Yan process. When 
the cross section of lepton pair production is considered varying with 
the  center-of-mass energy of the nucleon-nucleon collision, we find that 
the nuclear Drell-Yan(DY) ratio is suppressed due to the energy loss, which 
balances the overestimate of the DY ratio only in consideration of 
the  effect of nuclear parton distributions.

\end {abstract}

\section{Introduction}

In 1983, the European Muon Collaboration(EMC) \cite {EMC-1} 
surprisingly found that
the nucleon structure function, as  measured 
in deep-inelastic lepton-nucleus scattering
(DIS), varies with the target nucleus. This phenomenon has been known as 
the EMC effect.
Since the discovery of the EMC effect,
various models have been proposed to investigate the nuclear effect 
\cite {XRE,QR,Li1}. 
Among them, the $x-$rescaling \cite{XRE} or  $Q^2$-rescaling mechanism
\cite {QR} is commonly accepted to adquately explain the EMC effect. In addition, 
continuum dimuon production in high-energy hadron collisions, known as the
Drell-Yan(DY) process \cite{DY}, provides an independent 
measure of the modification of the quark structure of nuclei.
Recently,  many investigations  on the EMC and nuclear Drell-Yan effects
are still going on with great progress \cite{EMC-DY,YJJ}.
Several years ago, the E772 Collaboration \cite {Alde}
at Fermilab published the data of high-mass
dilepton production measured in the nuclear Drell-Yan (DY) process.
 These data aroused special attention in clarifying 
the various explanations of the nuclear effect  
on  the parton distributions.
Their results show that the ratio of DY dimuon yield per 
nucleon on a nuclear target to that on a free nucleon (from now on, shorten 
as the nuclear DY ratio) is slightly 
less than unity if the momentum fraction $x$ carried by 
a target quark is less than 0.1. The ratios over the range 
$0.10 \leq x \leq 0.30$, however, do not reveal distinct nuclear dependence.
Bickerstaff et al.
 \cite {Bickerstaff} found that although most of the theoretical
 models provide 
 good explanations for the EMC effect, they do not give a good description of 
the nuclear Drell-Yan ratio. Most of the theoretical models of the EMC 
effect overestimate the nuclear DY ratio. 

Several years ago, we put forward an extended $x-$rescaling model \cite {LGL} 
 in  which  different 
$x-$rescaling parameters  for the valence quarks and sea quarks (gluons) 
in the nucleon structure function are employed in consideration of  the   
nuclear momentum conservation. With a simultaneous consideration of the 
nuclear shadowing and nuclear momentum 
conservation, the experimental data of the EMC 
effect can be well explained in the whole x region.  However, similar to 
the prediction for nuclear DY ratio in  pion-excess and quark-cluster 
models \cite{Bickerstaff}, 
by using the obtained nuclear parton distributions in the extended 
$x$-rescaling model, the nuclear DY ratio is 
also overestimated. The difference of the  nuclear effects between 
the nuclear DY and  DIS  processes is not clear yet.
In this paper, we suggest an additional nuclear effect  due to 
 the energy loss in DY process. 
 We find that  the nuclear 
DY ratio is suppressed significantly  as a consequence of
continuous energy loss of the projectile nucleon to the target nucleon
in their successive binary nucleon-nucleon collisions. This suppression 
balances  the overestimate of the DY ratio only in consideration of the 
nuclear effect on the parton distributions. 
Therefore, a combination of these two types of nuclear effects can  give a good 
explanation of the experimental data of the nuclear DY ratio.

\section{Nuclear Parton Distributions in the Extended  $x-$rescaling Model}

To provide  the nuclear parton distributions which can be used to explain  
the experimental data of the EMC  effect  in the whole x region,
we work in our familiar extended $x-$rescaling model.
Let $I_{A(N)} (x, Q^2)$, I=V, S, G be 
the probability distributions of 
valence quarks(V), sea quarks(S), and gluons(G) in the nucleus A(or 
nucleon N), respectively. Then $K^I_{A(N)}(x,Q^2)=x I_{A(N)}(x,Q^2)$, 
I=V, S, G, are the  momentum distributions of valence quarks(V), sea 
quarks(S), and gluons(G) in the nucleus A (or nucleon N), respectively.
In Ref. \cite {LGL}, we pointed out that,  the nuclear binding 
effect together with  the  $x-$rescaling mechanism further introduced 
does not affect the valence quark number conservation. However, the nuclear 
momentum is no longer conserved in  the 
$x-$rescaling model. In order to keep nuclear 
momentum conservation, we extended  the $x-$rescaling 
model and employ different  $x-$rescaling 
parameters for the momentum 
distributions of valence quarks and sea quarks (gluons) in the nucleon 
structure function, i.e.,
\begin{equation}
 K^{V(S)}_{A} (x,Q^2)=K^{V(S)}_{N} (\delta_{V(S)}x,Q^2) ,
\end{equation}
Because the momentum distributions of sea quarks and gluons 
have similar forms, we take the same  $x-$rescaling parameter for 
the momentum distributions of sea quarks and gluons in the nucleus.
The numerical  result shows that, by properly choosing 
these parameters (one of them is determined according to the 
nuclear momentum conservation condition), one  can well explain 
the experimental data of the EMC effect. Because of its simple form and 
also for giving a good explanation of the EMC effect, the
extended $x-$rescaling model has been adopted by EMC \cite{EMCCITE} to 
fit their experimental data. 
Naturally, one hope that the nuclear DY ratio can also be well predicted
by using the obtained nuclear parton distributions. Unfortunately, the 
calculation results show that the nuclear DY ratio is overestimated only if
the nuclear effect on the parton distributions is considered(see the 
dotted-line in Fig. 1).
This indicates that  some other mechanisms of the nuclear effect  should 
be further taken into account.

\section{Energy Loss in Nuclear DY Process}

The Drell-Yan(DY) model \cite {DY} gives a good description of 
the continum of massive dimuon  pair production in the collision of 
proton with the nucleus $A$:
\begin{equation}
p +A \rightarrow \mu^+ \mu^- +X.
\end{equation}
This process is described as an electromagnetic 
annihilation of a quark (antiquark) in the proton $p$ and an 
antiquark (quark) in the nucleon embedded in the nucleus $A$ into a dimuon pair. 
The parton-model cross section for the DY process is given by 

\begin{eqnarray}
\frac{d^2\sigma}{dM^2dx_F}  =K  \frac{4 \pi \alpha^2}{9sM^2}
       \sum \limits_i e_i^2\frac{[q^p_i(x_1)
          \bar{q}^A_i(x_2)+\bar{q}^p_i(x_1)
          q^A_i(x_2)]}{\sqrt{x_F^2+4M^2/s}} , \label {DY1}
\end{eqnarray}
where $K$-factor, with $K \sim 2$, is due to next-to leading order QCD 
calculations \cite {QCDDY}, and  $\alpha$ is the 
fine-structure constant, $e_i$ is the fractional 
charge of the quark of flavor $i$, and $q_i^{p(A)}(x)$ and 
$\bar{q}^{p(A)}_i(x)$ are, respectively, the quark and anti-quark 
distributions in the proton(nucleon embedded in the nucleus $A$). 
The Feynman scaling 
variable $x_F$ is defined as
\begin{equation}
x_F=\frac{2p_l}{\sqrt{s}},
\end{equation}
where $\sqrt{s}$ is the nucleon-nucleon center-of-mass system(cms) 
energy and $p_l$ is
the longitudinal momentum of the virtual photon of mass $M$.
The quantities $x_{1,2}$ are related to $x_F$ and $M^2$ by
\begin{equation}
x_{1,2}=\frac {1}{2} (\sqrt{x_F^2+\frac{4M^2}{s}} \pm x_F),
\end{equation}
\begin{equation}
x_1-x_2=x_F,
\end{equation}
and
\begin{equation}
x_1 x_2=\frac {M^2}{s}.
\end{equation}
The above relations are easily extended to account for the evolution of 
quark structure functions with $Q^2$. 

Now let us  turn to the case in which the effect of energy loss in 
the initial states is taken into account.
For a nucleon-nucleus collision, the probability of having $n$ collisions at
an impact parameter $\vec{b}$ can be expressed as \cite {Thick}

\begin{equation}
P(\vec{b},n)=\frac{A!}{n!(A-n)!}[T(\vec{b})\sigma_{in}]^n [1-T(\vec{b})
\sigma_{in}]^{A-n},
\label {Pnb}
\end{equation}
where $\sigma_{in}$ ($\sim$ 30mb \cite {Thick}) is the non-diffractive cross 
section  for inelastic nucleon-nucleon collision, and  
$T(\vec{b})$ is the thickness function of the impact parameter $\vec{b}$.
The basic thickness function $T(\vec{b})$ can be well approximated by a Gaussian
function with a standard deviation $\beta_p$. If the collided nuclei are 
small($A \leq 32$),
their density function $\rho$ can also be taken to be a Gaussian function of
the spatial coordinates. Consequently, the thickness function can be
conveniently written as \cite {Thick}
\begin{equation}
T(\vec{b})=\displaystyle{exp} (-\vec{b}^2/2\beta_A^2)/2\pi\beta_A^2.
\end{equation}
In terms of the standard root-mean-squared-radius parameter $r_0^\prime$
for the nucleus A, the standard deviation $\beta_A$ is given by
\begin{equation}
\beta_A=r_0^\prime A ^{1/3}/\sqrt{3},
\end{equation}
here $r_0^\prime$ is found to be 1.05 fm in Ref. \cite {Thick},
and therefore 
\begin{equation}
\beta_A=0.606 A^{1/3}.
\end{equation}
For the nucleus with larger mass number $A(A > 32)$, the 
thickness function can be approximated by using a sharp-cutoff density 
distribution of the form \cite {Thick}
\begin{equation}
T(\vec{b})=\frac{3}{2 \pi R_A^3}\sqrt{R_A^2-\vec{b}^2}\theta (R_A-|\vec{b}|),
\end{equation}
where $R_A=r_0 A^{1/3}$ is the radius of colliding nucleus with $r_0=1.2$ fm.

In Eq. (\ref{Pnb}), the 
first factor on the right-hand side represents the number 
of combinations for finding $n$ collisions out of $A$ 
possible nucleon-nucleon encounters, the second factor gives 
the probability of exactly $n$ collisions
and the third factor gives the probability of having exactly $A-n$ misses. 
The total probability for the occurrence of an inelastic event in the 
collision of proton with the  nucleus $A$ at an impact parameter $\vec{b}$ is the sum of 
Eq. (\ref {Pnb}) from $n=1$ to $n=A$:

\begin{eqnarray}
\frac{d \sigma_{in}^{p-A}}{d \vec{b}} &=&\sum \limits_{n=1}^A P(n,\vec{b}) 
\nonumber \\
&=& 1-[1-T(\vec{b}) \sigma_{in}]^A.
\label {dsigmab}
\end{eqnarray}
Therefore, from Eq. (\ref {dsigmab}), the total inelastic cross section 
$\sigma _{in}^{p-A}$ for the collisions of protons with the nucleus $A$  is

\begin{equation}
\sigma_{in}^{p-A}=\int d \vec{b} \{1-[1-T(b) \sigma_{in}]^A \}.
\end{equation}
In an inelastic nucleon-nucleus collision without impact parameter selection, 
the number of nucleon-nucleon collisions $n$ (for $n=1$ to $A$)has a probability 
distribution $P(n)$. This is obtained by integrating $P(n,\vec{b})$ over 
all impact parameters:

\begin{equation}
P(n)=\frac{\int d \vec{b} P(n,\vec{b})}{\sum \limits_{n=1}^A \int d \vec{b}
P(n, \vec{b})}, \label {Pn}
\end{equation}
where the denominator is to ensure that $P(n)$ is properly normalized as
\begin{equation}
\sum\limits_{n=1}^A P(n)=1.
\end{equation}
From Eq. (\ref {dsigmab}), the  denominator of the 
right hand side of Eq. (\ref {Pn}) can be replaced by 
\begin{equation}
\sum \limits_{n=1}^A \int d \vec{b} P(n, \vec{b})=
\int d \vec{b}\{1-[1-T(\vec{b}) \sigma_{in}]^A \}.
\end{equation}
To describe the energy loss in the collision of  a proton with the nucleus  $A$, 
we start with remarks on the relative role of "soft" and "hard" interactions 
in nuclear collisions at very high energy: The incident proton interacts 
with spectator nucleon and makes soft (nonperturbative) minimum bias 
collisions before making the high $Q^2$ dimuon pair; During the
"soft" collisions, the projectile proton imparts energy to the struck
nucleon and therefore must loose energy;
Thus energy loss must affect the cross sections of 
producing dimuon pair.
After the projectile proton 
has  had additional $n$ collisions with 
nucleons embedded in the nucleus, the  cms energy of the colliding nucleons 
with  "hard" DY collisions   can be expressed as 

\begin{equation}
\sqrt{s^\prime}=\sqrt{s}-(n-1)\frac{d\sqrt{s}}{dn},
\end{equation}
where $\frac{d\sqrt{s}}{dn}$, generally taken as 0.2--0.4$GeV$,
is the cms energy loss per collision in the initial state.
Therefore, the cross section for the DY process can be re-written as

\begin{eqnarray}
\frac{d^2\sigma}{dM^2dx_F} = K \frac{\sqrt{s}}{\sqrt{s^\prime}} 
          \frac{4 \pi \alpha^2}{9s^\prime} 
          \sum\limits_i e_i^2
          \frac{[q^p_i(x^\prime_1)
          \bar{q}^A_i(x^\prime_2)+\bar{q}^p_i(x^\prime_1)
          q^A_i(x^\prime_2)]}{\sqrt{{x^\prime_F}^2+
          4M^2/s^\prime}} , \label {DY2}
\end{eqnarray}
where the rescaled quantities are defined as
\begin{equation}
x^\prime_F=\frac{2p_l}{\sqrt{s^\prime}}=r_s x_F,
\end{equation}
and 

\begin{equation}
x_{1,2}^\prime = r_s x_{1,2} , 
\end{equation}
with the cms energy ratio:
\begin{equation}
r_s = \frac{\sqrt{s}}{\sqrt{s^\prime}}.
\label {rs}
\end{equation}
The average cross section for the dimuon production in nuclear 
DY process can be expressed as
\begin{equation}
\langle \frac{d^2\sigma}{dM^2dx_F} \rangle=\sum\limits_{n=1}^A 
       P(n) \frac{d^2\sigma}{dM^2dx_F}.
\label {DY3}
\end{equation}
To make a comparison between the theoretical prediction for  the nuclear DY
ratio and the experimental data with respect to the variables $x_1$ and $x_2$, 
Eq. (\ref {DY2}) and Eq. (\ref {DY3}) can be re-expressed as
\begin{eqnarray}
\frac{d^2\sigma}{dx_1dx_2}  = K \frac{4 \pi \alpha^2}{9M^2} 
        \sum\limits_i e_i^2
    [q^p_i(r_s x_1)
          \bar{q}^A_i(r_sx_2)+\bar{q}^p_i(r_sx_1)
          q^A_i(r_sx_2)] , \label {DY4}
\end{eqnarray} 
and 

\begin{equation}
\langle \frac{d^2\sigma}{dx_1dx_2} \rangle=\sum\limits_{n=1}^A 
        P(n) \frac{d^2\sigma}{dx_1dx_2},
\end{equation}
respectively. It is noteworthy that $r_s$ in (\ref{rs}) is always  
greater than one if there exists an energy loss  in the collisions of  
 protons with the  nucleus $A$.

\section{Numerical Results}
To compare the theoretical prediction for nuclear Drell-Yan process with 
the experimental data of E772 collaboration \cite {Alde}, we introduce
the nuclear DY ratio as
\begin{equation}                              
T^{A/D}(x_2)=\frac{\int dx_1\langle \frac{d^2\sigma^{p-A}}{dx_1dx_2} \rangle}
      {\int dx_1 \frac{d^2\sigma^{p-D}}{dx_1dx_2}},
\label {DYR}
\end{equation}
where  $\frac{d^2\sigma^{p-D}}{dx_1dx_2}$ is the differential cross 
section  for 
the dimuon pair production in the proton-deuteron collision.   
The  integral range for $x_1$ in Eq.(\ref {DYR}) is determined according to
 the kinematic region  of  the
 experiment in  Ref. \cite {Alde}, i.e. 
 $x_1-x_2>0$, and $0.025 \leq x_2 \leq 0.30$.
By using  the free  parton distributions of GRV \cite {GRV} and taking  
the  $x-$rescaling parameters $\delta_V=1.026$ and $\delta_S=0.945$ for 
valence quarks and sea quarks, respectively, we calculate the nuclear 
Drell-Yan ratios for  $^{56}$Fe. In addition,
we assume that the gluon and the sea quark structure functions have similar
shadowing effects at small $x_2$,  and according to Ref. \cite{RSH}, we 
introduce the shadowing factor $R^A_{sh}$ both for gluons and sea quarks as
$$ 
   R^A_{sh}(x_2)=\left\{
      \begin{array}{ll}
   1+alnAln(x_2/0.08), & x_2 < 0.08,\\
   1+blnAln(x_2/0.08)ln(x_2/0.24), & 0.08<x_2<0.3
      \end{array}
\right.
$$
where the parameters a, b are taken as 0.025 and -0.02, respectively.
The calculation results with $\frac{d \sqrt{s}}{d n}=0.0, 0.2, 0.4 GeV$ shown in 
Fig. 1 indicate that the energy 
loss effect is essential in the  
explanation  of the nuclear DY ratio.

\section{Discussion and Summary}
To sum up, by means of the nuclear parton distributions in the 
extended $x-$rescaling model on which the experimental data of the EMC effect
can be well explained, we investigate the  nuclear 
DY process focusing on the continuous energy loss of the projectile 
nucleon to target nucleons in their successive nucleon-nucleon
collisions. We find that the nuclear DY ratio will be  overestimated
if the nuclear effect on the parton distributions is the only factor 
considered. The calculation results show that the nuclear DY ratio is 
sensitive to the change of  sea  quark distributions. On one hand, to 
adequately explain the antishadowing effect and the EMC effect, there 
should exist an enhancement of the sea quarks in the nuclear quark distributions 
in the range of  $0.1 \le x \le 0.3$,  resulting in the overestimation 
of the nuclear DY ratio; On the other hand, the energy loss
causes a suppression of the nuclear DY ratio, which balances the 
overestimate of the nuclear DY ratio due to the nuclear effect 
on the parton distributions. So, in the nuclear DY process, 
there is no distinct  nuclear effect  as observed in DIS process  due to 
the the combination of the nuclear effect
on the parton distributions and the energy loss effect.   
Similarly, the J/$\psi$ suppression is also partially due to  the 
energy loss in the initial states.

\begin{center}
{\bf Acknowlegements}                                       
\end{center}
One of the authors (J. J. Yang) would like to thank Prof. W. Q. Chao and 
Dr. A. Tai of IHEP  for very helpful discussion. This work was 
supported in part by the Natural Science Foundation of China 
grant No. 19775051 and the Natural  Science Foundation of Jiangsu 
Province, China.

\centerline{\bf \huge Figure Caption} 

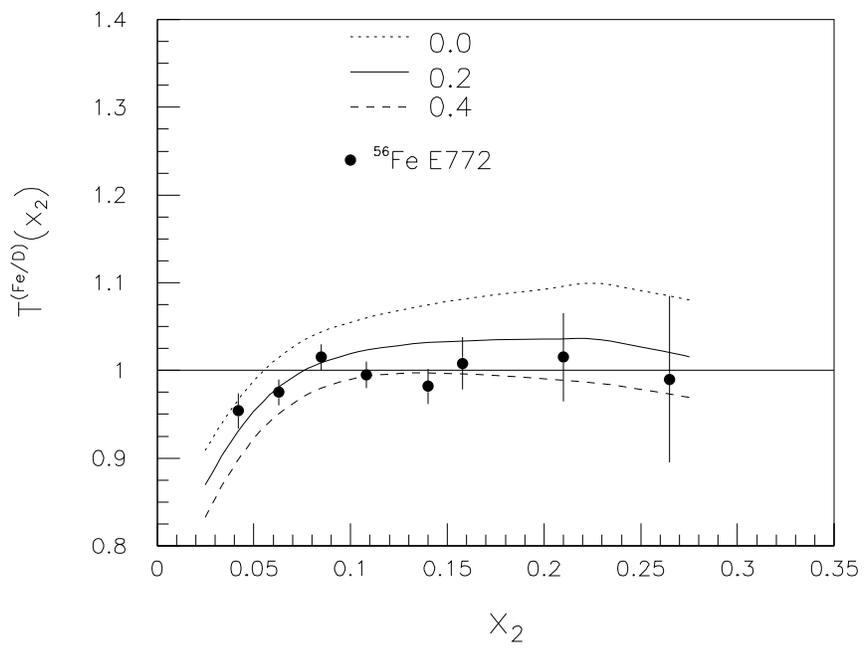
Fig. 1. The nuclear Drell-Yan ratios $T^{Fe/D}(x_2)$ predicted with
the energy loss ( solid  curve for $\frac{d\sqrt{s}}{dn}$=0.2 GeV,  
dashed  curve for $\frac{d\sqrt{s}}{dn}$=0.4 GeV) and without the energy 
loss(dotted  curve  for $\frac{d\sqrt{s}}{dn}$=0.0 GeV).
The experimental data are taken from the E772 Collaboration \cite {Alde}.

\end{document}